\documentclass[sigconf,anonymous=False]{acmart}

\usepackage{booktabs} % For formal tables
\usepackage[english]{babel}
\usepackage{subcaption}
\usepackage{tabularx}
\usepackage[inline]{enumitem}
\usepackage{multirow}

\setcopyright{acmlicensed}
\acmConference[LND4IR '18]{ACM SIGIR 2018 Workshop on Learning from Limited or Noisy Data}{July 12, 2018}{Ann Arbor, Michigan, USA} 
\acmYear{2018}
\copyrightyear{2018}
\acmDOI{}
\acmISBN{}
\acmPrice{}

\begin{document}

% % DOI
% \copyrightyear{2018} 
% \acmYear{2018} 
% \setcopyright{acmcopyright}
% \acmConference[SIGIR '18]{The 41st International ACM SIGIR Conference on Research & Development in Information Retrieval}{July 8--12, 2018}{Ann Arbor, MI, USA}
% % \acmBooktitle{SIGIR '18: The 41st International ACM SIGIR Conference on Research & Development in Information Retrieval, July 8--12, 2018, Ann Arbor, MI, USA}
% \acmPrice{15.00}
% \acmDOI{10.1145/3209978.3210118}
% \acmISBN{978-1-4503-5657-2/18/07}

\title[]{Distributed Evaluations: Ending Neural Point Metrics}
% LETOR: A benchmark collection for research on learning to rank for information retrieval

\author{Daniel Cohen \quad Scott Jordan \quad W. Bruce Croft}
\affiliation{%
	\institution{
	 University of Massachusetts Amherst, Amherst, MA, USA}
	%\streetaddress{P.O. Box 1212}
	%\city{Dublin} 
	%\state{Ohio} 
	%\postcode{43017-6221}
}
\email{{dcohen, sjordan, croft}@cs.umass.edu}

\renewcommand{\shortauthors}{D. Cohen et al.}

\begin{abstract}
\noindent With the rise of neural models across the field of information retrieval, numerous publications have incrementally pushed the envelope of performance for a multitude of IR tasks. However, these networks often sample data in random order, are initialized randomly, and their success is determined by a single evaluation score. 
These issues are aggravated by neural models achieving incremental improvements from previous neural baselines, leading to multiple near state of the art models that are difficult to reproduce and quickly become deprecated. As neural methods are starting to be incorporated into low resource and noisy collections that further exacerbate this issue, we propose evaluating neural models both over multiple random seeds and a set of hyperparameters within $\epsilon$ distance of the chosen configuration for a given metric.

%\ylcomment{The released data set consisted of thousands of questions with annotated answer passages provides an openly available test bed for future research on non-factoid answer passage retrieval. }
\end{abstract}

% \keywords{deep learning, evaluation}

\maketitle

%\input{intro}
%Outline and key points 
%1. High quality open available benchmark data sets are critical for research progress on various IR tasks. For instance, ImageNet, TREC GOV2, CLUE Web, SQUAD, etc...
%2. In terms of question answering, there are some existing benchmark data sets including TREC QA, Wiki QA, Insurance QA... But most of these data sets are on factoid question answering and the answers are short snippets or sentences on answer facts. There is a lack of high quality non-factoid question answering data sets.
%3. To address this issue, WebAP was released as an initial effort for ... But the number of questions of WebAP is not enough to train powerful deep learning models. There are no previous large non-factoid QA data sets with thousands of questions...
%4. Thus in this paper, we propose ...

\section{Introduction}
As neural methods have become some of the most effective models for learning representations where traditional hand crafted features have failed to perform~\cite{moschitti,mitra2017learning,Guo-DRMM}, there has been a large increase in publications using these approaches. This has allowed the field to move from handcrafting features to handcrafting larger architectures that can learn relevance with millions of parameters. While this approach has made significant strides in the field of IR, reproducible results have become a significant concern within the community~\cite{ecirDur18}. Often, these state of the art results cannot be replicated due to a small issue such as batch size, data preprocessing, random seed, or other hyperparameters of the model. While Choromanska et al.~\cite{lossSurfaceChoromanska} have demonstrated that local minimas are sufficiently close to the global minimum, this is not calibrated with local minimas being a sufficient in evaluation space such as mean precision or recall~\cite{NIPS2012_4646}; a model that achieves a similar loss value is therefore not calibrated to a similar ranking score.

Thus, we propose addressing this issue by introducing a new evaluation method for neural retrieval. Rather than pointwise comparisons of single scores, models would be reported with a probability density function over random seeds. This would allow future work to not only compare the mean performance score, but to examine the sensitivity of new architectures or training methods. Past work~\cite{cart1} has evaluated the efficacy of traditional statistical methods for system performance, but don't take into account measuring the similarity with respect to training or hyperparameter selection.  

\section{Volatility of Neural Models}

\begin{table}[h]
\centering
 \caption{Sensitivity of MAP across CQA and WikiQA collections over multiple random seeds}
 \begin{tabular}{lcc}
 \toprule

 Method &   CQA & WikiQA\\
 \midrule
 LSTM         & $.665\pm.004$ & $.592\pm.021$\\
 Multitask LSTM  & $.615\pm.009$ & $.572\pm.060$\\
 
 \bottomrule
 \end{tabular}%
 \label{tab:IRmetrics}
 \vspace{-0.3cm}
\end{table}

\begin{figure}[htp]
\centering
\includegraphics[width=1\linewidth]{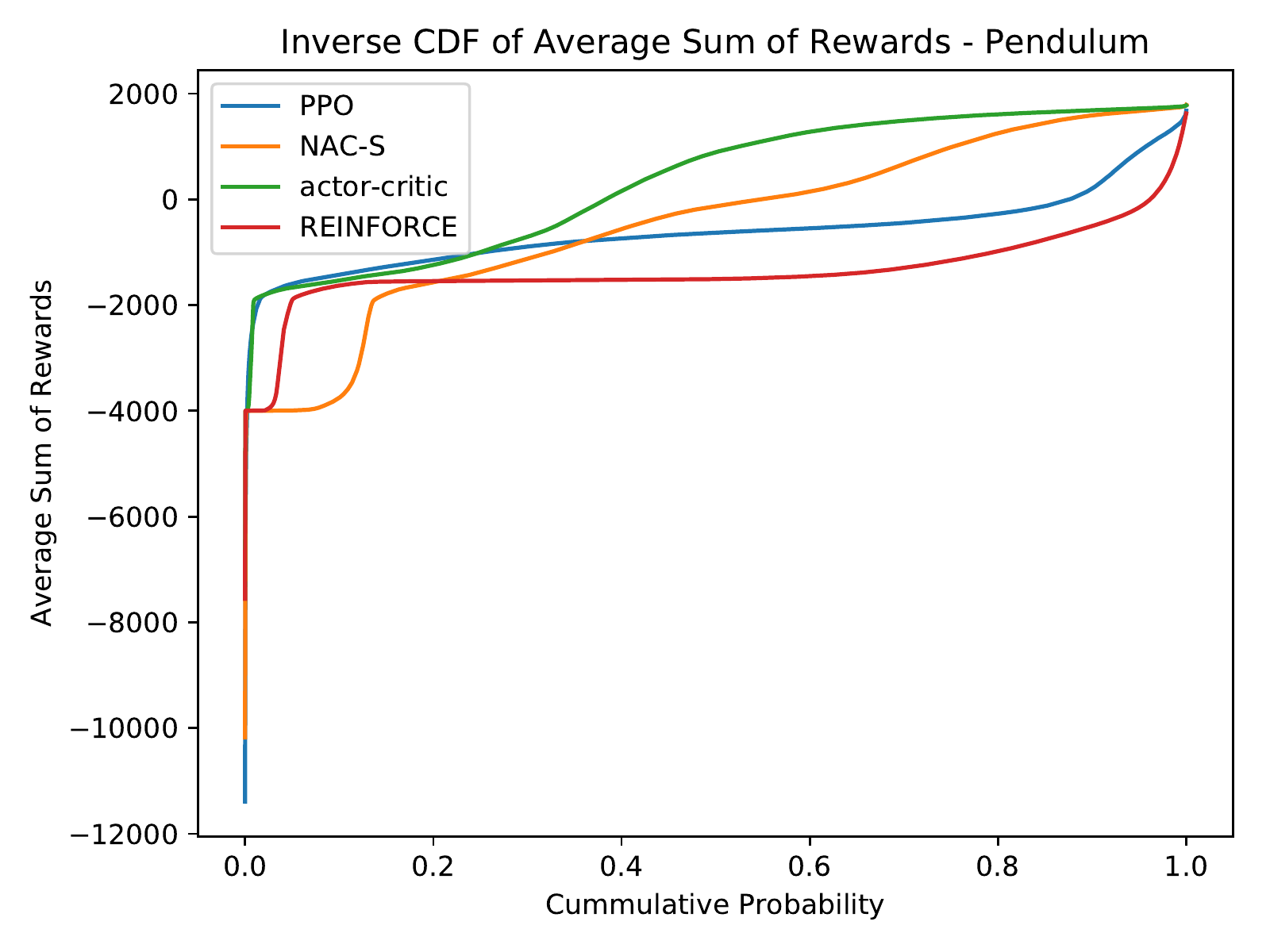}
\caption{Full performance distribution of on the pendulum swing and balance task}
\label{fig:rl}
\end{figure}

As D\"{u}r et al.~\cite{ecirDur18} have demonstrated, state of the art neural models are extremely susceptible to small changes in hyperparameters, initialization and even random seeds. As IR neural models are often trained and evaluated over a limited number of training queries, this variance is not uncommon. To exemplify this, we conduct a small experiment over multiple random seeds by evaluating a conventional short text retrieval architecture~\cite{cohen-ictir} compared with the same model with an additional multitask component to predict part of speech information~\cite{long2015aMulti}. This experiment was conducted over two collections. \textit{CQA} which is the combination of nfl6~\cite{cohen-ictir} and Yahoo's \textit{manner} collection commonly referred to as L4~\cite{yahoo}. This combined collection has close to 200,000 individual queries. The other is WikiQA~\cite{wikiQA}, which consists of approximately 2000 training queries.\\
As seen in Table~\ref{tab:IRmetrics}, the large amount of data available within the CQA collection to evaluate these two methods results in a relatively stable performance across random seeds. However, moving to a lower resource collection results in a much higher variance across initialization. Of particular interest is that Multitask LSTM could be portrayed as the superior model under a certain set of random initial conditions. \\
As recent work has started using reinforcement learning (RL) to handle noisy approaches~\cite{irgan}, the importance of fully documenting a proposed model's performance becomes an even greater issue. 
The REINFORCE algorithm \cite{williams92}, used in \cite{irgan} is known to have exceptionally high variance in the gradient estimates, which translates to high variance in the performance metrics. 
%High variance gradient estimates are a common problem in a family of RL algorithms known as policy gradient methods, of which REINFORCE belongs. 
%
To demonstrate the importance of using distributed evaluations, we implement several RL algorithms that have been shown empirically to be more stable than the one used in IRGAN~\cite{irgan,thomas_bias_2014,schulman_ppo_2017,SuttonBarto}. 
However, even with these new algorithms, the stochastic optimization process has high variance and has led to issues with reproducibility~\cite{henderson_drlmatters_2018}. 
Any IR model using reinforcement learning needs to be evaluated over many trials to accurately convey the results. 
As seen in Figure~\ref{fig:rl}, we show the sensitivity of four reinforcement learning algorithms: REINFORCE~\cite{williams92}, actor-critic~\cite{SuttonBarto}, NAC-s \cite{thomas_bias_2014}, and PPO \cite{schulman_ppo_2017} on a common benchmark, the pendulum swing up task~\cite{doya2000reinforcement}. 
This task involves suspending a pendulum inverted through single directional inputs. While this does not seem similar to IR, the relative simplicity of this task exemplifies the inherent issues with trying to use RL approaches on more complicated spaces without proper evaluation. The performance is the average sum of reward the RL agent sees over its lifetime. We plot the inverse CDF of agents performance after running 125 thousand random settings of the hyper-parameters and random seeds.

%scott note: plot says RETURNS not sum of rewards. Typo in last sentence before the plot. 

\section{Distributed Evaluations}

To circumvent the issues mentioned in the previous section, we propose a two fold evaluation approach to neural models. First, final evaluation scores should be conducted over multiple random seeds. This creates a distribution of scores, and provides an illustration of the sensitivity of the proposed model to noise. Second, a subsequent set of scores would be evaluated over a small $\epsilon$-ball of the top hyperparameters of the best performing model. The impact of small changes in the hyperparameter space reveals the robustness of the model over the small perturbations to architecture choices.
\\
% Improved Comparison to Baselines}
Using these two approaches, it now becomes viable to create a smoothed distribution of scores from a model and evaluate a novel architecture with the additional information. Using KL-divergence, $KL(P\|Q) = \int_{-\infty}^{\infty} p(x) \log \frac{p(x)}{q(x)}dx$, one can not only examine point statistics such as mean and variance, but also the similarity of each model's sensitivity to randomness and hyperparameters.

\section{Conclusion}
In this paper, we address the issue of under-reporting the performance of models that are highly susceptible to noise both in the training data, but also within the model itself. While the proposed distributed evaluation requires greater computation than taking the result of a single run, hyperparameter tuning within a small convex hull is common practice when fine-tuning a model for a collection. Thus one need only include these results in the final paper and not incur additional overhead. \\
With the recent push to release code for the public, setting a standard of distributed results would bring the field one step closer to allowing these methods to be reproducible.

\section{Acknowledgements}
This work was supported in part by the Center for Intelligent Information Retrieval. Any opinions, findings and conclusions 
or recommendations expressed in this material are those of the authors and do not necessarily reflect those of the sponsor.

 %\ylcomment{We present this new collection for non-factoid answer passage retrieval with various benchmark results so that others can extend our research on this challenging task in the future.}
 %In summary, we present this collection in the segmented window format used for evaluation in addition to the raw start and end indices within the Wikipedia text document for the IR community.
%\ylcomment{The Section 5 is a bit short.}

%\input{related}
%\input{data}
%\input{method}
%\input{exp}
%\input{conclusion}
% \enlargethispage{-2\baselineskip}% pnb: fix widowed section header

\bibliographystyle{ACM-Reference-Format}
 \bibliography{reference}  % sigproc.bib is the name of the Bibliography in this case

%%% -*-BibTeX-*-
%%% Do NOT edit. File created by BibTeX with style
%%% ACM-Reference-Format-Journals [18-Jan-2012].

\begin{thebibliography}{00}

%%% ====================================================================
%%% NOTE TO THE USER: you can override these defaults by providing
%%% customized versions of any of these macros before the \bibliography
%%% command.  Each of them MUST provide its own final punctuation,
%%% except for \shownote{}, \showDOI{}, and \showURL{}.  The latter two
%%% do not use final punctuation, in order to avoid confusing it with
%%% the Web address.
%%%
%%% To suppress output of a particular field, define its macro to expand
%%% to an empty string, or better, \unskip, like this:
%%%
%%% \newcommand{\showDOI}[1]{\unskip}   % LaTeX syntax
%%%
%%% \def \showDOI #1{\unskip}           % plain TeX syntax
%%%
%%% ====================================================================

\ifx \showCODEN    \undefined \def \showCODEN     #1{\unskip}     \fi
\ifx \showDOI      \undefined \def \showDOI       #1{#1}\fi
\ifx \showISBNx    \undefined \def \showISBNx     #1{\unskip}     \fi
\ifx \showISBNxiii \undefined \def \showISBNxiii  #1{\unskip}     \fi
\ifx \showISSN     \undefined \def \showISSN      #1{\unskip}     \fi
\ifx \showLCCN     \undefined \def \showLCCN      #1{\unskip}     \fi
\ifx \shownote     \undefined \def \shownote      #1{#1}          \fi
\ifx \showarticletitle \undefined \def \showarticletitle #1{#1}   \fi
\ifx \showURL      \undefined \def \showURL       {\relax}        \fi
% The following commands are used for tagged output and should be
% invisible to TeX
\providecommand\bibfield[2]{#2}
\providecommand\bibinfo[2]{#2}
\providecommand\natexlab[1]{#1}
\providecommand\showeprint[2][]{arXiv:#2}

\bibitem[\protect\citeauthoryear{Calauz\`{e}nes, Usunier, and
  Gallinari}{Calauz\`{e}nes et~al\mbox{.}}{2012}]%
        {NIPS2012_4646}
\bibfield{author}{\bibinfo{person}{Cl\'{e}ment Calauz\`{e}nes},
  \bibinfo{person}{Nicolas Usunier}, {and} \bibinfo{person}{Patrick
  Gallinari}.} \bibinfo{year}{2012}\natexlab{}.
\newblock \showarticletitle{On the (Non-)existence of Convex, Calibrated
  Surrogate Losses for Ranking}. In \bibinfo{booktitle}{{\em NIPS}},
  \bibfield{editor}{\bibinfo{person}{F.~Pereira}, \bibinfo{person}{C.~J.~C.
  Burges}, \bibinfo{person}{L.~Bottou}, {and} \bibinfo{person}{K.~Q.
  Weinberger}} (Eds.). \bibinfo{pages}{197--205}.
\newblock


\bibitem[\protect\citeauthoryear{Carterette}{Carterette}{2012}]%
        {cart1}
\bibfield{author}{\bibinfo{person}{Benjamin~A. Carterette}.}
  \bibinfo{year}{2012}\natexlab{}.
\newblock \showarticletitle{Multiple Testing in Statistical Analysis of
  Systems-based Information Retrieval Experiments}.
\newblock \bibinfo{journal}{{\em ACM Trans. Inf. Syst.\/}}
  \bibinfo{volume}{30}, \bibinfo{number}{1}, Article \bibinfo{articleno}{4}
  (\bibinfo{date}{March} \bibinfo{year}{2012}), \bibinfo{numpages}{34}~pages.
\newblock
\showISSN{1046-8188}
\showDOI{%
\url{https://doi.org/10.1145/2094072.2094076}}


\bibitem[\protect\citeauthoryear{Choromanska, Henaff, Mathieu, Arous, and
  LeCun}{Choromanska et~al\mbox{.}}{2014}]%
        {lossSurfaceChoromanska}
\bibfield{author}{\bibinfo{person}{Anna Choromanska}, \bibinfo{person}{Mikael
  Henaff}, \bibinfo{person}{Micha{\"{e}}l Mathieu},
  \bibinfo{person}{G{\'{e}}rard~Ben Arous}, {and} \bibinfo{person}{Yann
  LeCun}.} \bibinfo{year}{2014}\natexlab{}.
\newblock \showarticletitle{The Loss Surface of Multilayer Networks}.
\newblock \bibinfo{journal}{{\em CoRR\/}}  \bibinfo{volume}{abs/1412.0233}
  (\bibinfo{year}{2014}).
\newblock
\showeprint[arxiv]{1412.0233}
\showURL{%
\url{http://arxiv.org/abs/1412.0233}}


\bibitem[\protect\citeauthoryear{Cohen and Croft}{Cohen and Croft}{[n. d.]}]%
        {cohen-ictir}
\bibfield{author}{\bibinfo{person}{Daniel Cohen} {and}
  \bibinfo{person}{W.~Bruce Croft}.} \bibinfo{year}{[n. d.]}\natexlab{}.
\newblock \showarticletitle{End to End Long Short Term Memory Networks for
  Non-Factoid Question Answering}. In \bibinfo{booktitle}{{\em ICTIR '16}}.
\newblock


\bibitem[\protect\citeauthoryear{Doya}{Doya}{2000}]%
        {doya2000reinforcement}
\bibfield{author}{\bibinfo{person}{Kenji Doya}.}
  \bibinfo{year}{2000}\natexlab{}.
\newblock \showarticletitle{Reinforcement learning in continuous time and
  space}.
\newblock \bibinfo{journal}{{\em Neural computation\/}} \bibinfo{volume}{12},
  \bibinfo{number}{1} (\bibinfo{year}{2000}), \bibinfo{pages}{219--245}.
\newblock


\bibitem[\protect\citeauthoryear{D{\"{u}}r, Rauber, and Filzmoser}{D{\"{u}}r
  et~al\mbox{.}}{2018}]%
        {ecirDur18}
\bibfield{author}{\bibinfo{person}{Alexander D{\"{u}}r},
  \bibinfo{person}{Andreas Rauber}, {and} \bibinfo{person}{Peter Filzmoser}.}
  \bibinfo{year}{2018}\natexlab{}.
\newblock \showarticletitle{Reproducing a Neural Question Answering
  Architecture Applied to the SQuAD Benchmark Dataset: Challenges and Lessons
  Learned}. In \bibinfo{booktitle}{{\em {ECIR} 2018, Grenoble, France, March
  26-29, 2018}}. \bibinfo{pages}{102--113}.
\newblock
\showDOI{%
\url{https://doi.org/10.1007/978-3-319-76941-7_8}}


\bibitem[\protect\citeauthoryear{Guo, Fan, Ai, and Croft}{Guo
  et~al\mbox{.}}{2016}]%
        {Guo-DRMM}
\bibfield{author}{\bibinfo{person}{Jiafeng Guo}, \bibinfo{person}{Yixing Fan},
  \bibinfo{person}{Qingyao Ai}, {and} \bibinfo{person}{W.~Bruce Croft}.}
  \bibinfo{year}{2016}\natexlab{}.
\newblock \showarticletitle{A Deep Relevance Matching Model for Ad-hoc
  Retrieval}. In \bibinfo{booktitle}{{\em CIKM '16}}. \bibinfo{publisher}{ACM},
  \bibinfo{address}{New York, NY, USA}, \bibinfo{pages}{55--64}.
\newblock
\showISBNx{978-1-4503-4073-1}
\showDOI{%
\url{https://doi.org/10.1145/2983323.2983769}}


\bibitem[\protect\citeauthoryear{Henderson, Islam, Bachman, Pineau, Precup, and
  Meger}{Henderson et~al\mbox{.}}{[n. d.]}]%
        {henderson_drlmatters_2018}
\bibfield{author}{\bibinfo{person}{Peter Henderson}, \bibinfo{person}{Riashat
  Islam}, \bibinfo{person}{Philip Bachman}, \bibinfo{person}{Joelle Pineau},
  \bibinfo{person}{Doina Precup}, {and} \bibinfo{person}{David Meger}.}
  \bibinfo{year}{[n. d.]}\natexlab{}.
\newblock \showarticletitle{Deep Reinforcement Learning That Matters}. In
  \bibinfo{booktitle}{{\em {AAAI}, New Orleans, Louisiana, USA, February 2-7,
  2018}} (2018).
\newblock
\showURL{%
\url{https://www.aaai.org/ocs/index.php/AAAI/AAAI18/paper/view/16669}}


\bibitem[\protect\citeauthoryear{Long and Wang}{Long and Wang}{2015}]%
        {long2015aMulti}
\bibfield{author}{\bibinfo{person}{Mingsheng Long} {and}
  \bibinfo{person}{Jianmin Wang}.} \bibinfo{year}{2015}\natexlab{}.
\newblock \showarticletitle{Learning Multiple Tasks with Deep Relationship
  Networks}.
\newblock \bibinfo{journal}{{\em CoRR\/}}  \bibinfo{volume}{abs/1506.02117}
  (\bibinfo{year}{2015}).
\newblock
\showeprint[arxiv]{1506.02117}
\showURL{%
\url{http://arxiv.org/abs/1506.02117}}


\bibitem[\protect\citeauthoryear{Mitra, Diaz, and Craswell}{Mitra
  et~al\mbox{.}}{2017}]%
        {mitra2017learning}
\bibfield{author}{\bibinfo{person}{Bhaskar Mitra}, \bibinfo{person}{Fernando
  Diaz}, {and} \bibinfo{person}{Nick Craswell}.}
  \bibinfo{year}{2017}\natexlab{}.
\newblock \showarticletitle{Learning to match using local and distributed
  representations of text for web search}. In \bibinfo{booktitle}{{\em WWW
  17}}. \bibinfo{pages}{1291--1299}.
\newblock


\bibitem[\protect\citeauthoryear{Schulman, Wolski, Dhariwal, Radford, and
  Klimov}{Schulman et~al\mbox{.}}{2017}]%
        {schulman_ppo_2017}
\bibfield{author}{\bibinfo{person}{John Schulman}, \bibinfo{person}{Filip
  Wolski}, \bibinfo{person}{Prafulla Dhariwal}, \bibinfo{person}{Alec Radford},
  {and} \bibinfo{person}{Oleg Klimov}.} \bibinfo{year}{2017}\natexlab{}.
\newblock \showarticletitle{Proximal Policy Optimization Algorithms}.
\newblock \bibinfo{journal}{{\em CoRR\/}}  \bibinfo{volume}{abs/1707.06347}
  (\bibinfo{year}{2017}).
\newblock
\showeprint[arxiv]{1707.06347}
\showURL{%
\url{http://arxiv.org/abs/1707.06347}}


\bibitem[\protect\citeauthoryear{Severyn and Moschitti}{Severyn and
  Moschitti}{2015}]%
        {moschitti}
\bibfield{author}{\bibinfo{person}{Aliaksei Severyn} {and}
  \bibinfo{person}{Alessandro Moschitti}.} \bibinfo{year}{2015}\natexlab{}.
\newblock \showarticletitle{Learning to Rank Short Text Pairs with
  Convolutional Deep Neural Networks}. In \bibinfo{booktitle}{{\em SIGIR}} {\em
  (\bibinfo{series}{SIGIR '15})}. \bibinfo{publisher}{ACM},
  \bibinfo{address}{New York, NY, USA}, \bibinfo{pages}{373--382}.
\newblock
\showISBNx{978-1-4503-3621-5}
\showDOI{%
\url{https://doi.org/10.1145/2766462.2767738}}


\bibitem[\protect\citeauthoryear{Surdeanu, Ciaramita, and Zaragoza}{Surdeanu
  et~al\mbox{.}}{2008}]%
        {yahoo}
\bibfield{author}{\bibinfo{person}{Mihai Surdeanu},
  \bibinfo{person}{Massimiliano Ciaramita}, {and} \bibinfo{person}{Hugo
  Zaragoza}.} \bibinfo{year}{2008}\natexlab{}.
\newblock \showarticletitle{Learning to rank answers on large online QA
  collections}. In \bibinfo{booktitle}{{\em ACL:HLT}}.
  \bibinfo{pages}{719--727}.
\newblock


\bibitem[\protect\citeauthoryear{Sutton and Barto}{Sutton and Barto}{1998}]%
        {SuttonBarto}
\bibfield{author}{\bibinfo{person}{R.~S. Sutton} {and} \bibinfo{person}{A.~G.
  Barto}.} \bibinfo{year}{1998}\natexlab{}.
\newblock \bibinfo{booktitle}{{\em Reinforcement Learning: {A}n Introduction}}.
\newblock \bibinfo{publisher}{MIT Press}, \bibinfo{address}{Cambridge, MA}.
\newblock


\bibitem[\protect\citeauthoryear{Thomas}{Thomas}{2014}]%
        {thomas_bias_2014}
\bibfield{author}{\bibinfo{person}{Philip Thomas}.}
  \bibinfo{year}{2014}\natexlab{}.
\newblock \showarticletitle{Bias in Natural Actor-Critic Algorithms}. In
  \bibinfo{booktitle}{{\em {ICML} 2014, Beijing, China, 21-26 June 2014}}.
  \bibinfo{pages}{441--448}.
\newblock
\showURL{%
\url{http://jmlr.org/proceedings/papers/v32/thomas14.html}}


\bibitem[\protect\citeauthoryear{Wang, Yu, Zhang, Gong, Xu, Wang, Zhang, and
  Zhang}{Wang et~al\mbox{.}}{2017}]%
        {irgan}
\bibfield{author}{\bibinfo{person}{Jun Wang}, \bibinfo{person}{Lantao Yu},
  \bibinfo{person}{Weinan Zhang}, \bibinfo{person}{Yu Gong},
  \bibinfo{person}{Yinghui Xu}, \bibinfo{person}{Benyou Wang},
  \bibinfo{person}{Peng Zhang}, {and} \bibinfo{person}{Dell Zhang}.}
  \bibinfo{year}{2017}\natexlab{}.
\newblock \showarticletitle{{IRGAN:} {A} Minimax Game for Unifying Generative
  and Discriminative Information Retrieval Models}. In \bibinfo{booktitle}{{\em
  {ACM} {SIGIR}, Shinjuku, Tokyo, Japan, August 7-11, 2017}}.
  \bibinfo{pages}{515--524}.
\newblock
\showDOI{%
\url{https://doi.org/10.1145/3077136.3080786}}


\bibitem[\protect\citeauthoryear{Williams}{Williams}{1992}]%
        {williams92}
\bibfield{author}{\bibinfo{person}{Ronald~J. Williams}.}
  \bibinfo{year}{1992}\natexlab{}.
\newblock \showarticletitle{Simple Statistical Gradient-Following Algorithms
  for Connectionist Reinforcement Learning}.
\newblock \bibinfo{journal}{{\em Machine Learning\/}}  \bibinfo{volume}{8}
  (\bibinfo{year}{1992}), \bibinfo{pages}{229--256}.
\newblock
\showDOI{%
\url{https://doi.org/10.1007/BF00992696}}


\bibitem[\protect\citeauthoryear{Yang, Yih, and Meek}{Yang
  et~al\mbox{.}}{2015}]%
        {wikiQA}
\bibfield{author}{\bibinfo{person}{Yi Yang}, \bibinfo{person}{Wen{-}tau Yih},
  {and} \bibinfo{person}{Christopher Meek}.} \bibinfo{year}{2015}\natexlab{}.
\newblock \showarticletitle{WikiQA: {A} Challenge Dataset for Open-Domain
  Question Answering}. In \bibinfo{booktitle}{{\em {EMNLP} 2015, Lisbon,
  Portugal, September 17-21, 2015}}. \bibinfo{pages}{2013--2018}.
\newblock
\showURL{%
\url{http://aclweb.org/anthology/D/D15/D15-1237.pdf}}


\end{thebibliography}

\end{document}